\documentclass[12pt,openany,a4paper]{article}

\usepackage[english]{babel}

\usepackage{cmap}  \usepackage[T2A]{fontenc}
\usepackage[pdftex,unicode=true,bookmarks=true]{hyperref}

\usepackage{xr}
\usepackage{indentfirst}

\usepackage{latexsym}
\usepackage{amsfonts}
\usepackage{amsmath}
\usepackage{amssymb}

\usepackage{wasysym}

\usepackage{cancel}

\newcommand{\be}{\begin{equation}}
\newcommand{\ee}{\end{equation}}
\newcommand{\bean}{\begin{eqnarray*}}
\newcommand{\eean}{\end{eqnarray*}}

\newcommand{\bea}{\begin{eqnarray}}
\newcommand{\eea}{\end{eqnarray}}

\newcommand{\const}{{\rm const}}

\newcommand{\Div}{{\rm div\,}}
\newcommand{\grad}{{\rm grad\,}}

\newcommand{\E}{\mathrm{e}}
\newcommand{\I}{\mathrm{i}}

\newcommand{\p}{\partial}

\title{Domain wall nonlinear quantization}
\author{M.~G.~Ivanov\thanks{\href{mailto:ivanov.mg@mipt.ru}{ivanov.mg@mipt.ru}}\\
\small Moscow Institute of Physics and Technology\\
\small Department of Theoretical Physics; Laboratory of Quantum Information Theory\\
\small 9 Institutskiy per., Dolgoprudny, Moscow Region, 141701, Russian Federation}
\date{June 8, 2020}

\begin{document}

\maketitle

\begin{abstract}
The nonlinear quantization of the domain wall (relativistic membrane of codimension 1) is considered.
The membrane dust equation is considered as an analogue of the Hamilton-Jacobi equation, which allows us to construct its quantum analogue. 
The resulting equation has the form of a nonlinear Klein-Fock-Gordon equation. 
It can be interpreted as the mean field approximation for a quantum domain wall.
Dispersion relations are obtained for small perturbations (in a linear approximation).
The group speed of perturbations does not exceed the speed of light.
For perturbations propagating along the domain wall, in addition to the massless mode (as in the classical case), a massive one appears.
The result may be interesting in condensed matter theory and in membrane quantization in superstring and supergravity theories.
\end{abstract}

\tableofcontents
\thispagestyle{empty}

\newpage

\section{Introduction}

The quantization of extended objects is of significant interest in the program of geometrization of physics. The most significant advances in this field are considered to be the quantization of the boson string and superstring. String theory has naturally led to the consideration of membranes of other dimensions, for which the quantization problem is also of great interest \cite{obzor}.

Strings and membranes are degenerate continuous media, so quantization can be interesting for condensed matter physics too \cite{perturb}.

We develop an approach based on the Hamilton-Jacobi equation.
Partially similar approaches was used in a series of earlier papers (see \cite{hosotani1999} and references wherein).
The main difference is the way the world surface is parameterized to quantize.
The paper \cite{hosotani1999} used scalar fields with gradients tangent to the world surface of the membrane.
The resulting quntized equations was essentially nonlinear, having no linear term.
We use a scalar field with a gradient orthogonal to the world surface of the membrane.
In the case of codimension 1 (domain wall), the resulting equation is the nonlinear Klein-Fock-Gordon equation.
It is nonlinear, but includes standard linear terms.

In this paper, we develop the following scheme for the membrane quantization.
\begin{enumerate}
\item  Transition from a single membrane to a continuous distribution of membranes (the membrane dust).
\item  Description of continuous membrane dust using the Hamilton-Jacobi type equation.
\item  Reconstruction of the classical generalized Hamiltonian from the Hamilton-Jacobi type equation.
\item  Replacing the classical generalized Hamiltonian with a quantum one.
\end{enumerate}

This scheme can be considered as a generalization of canonical quantization.
In particular, if you start the procedure with the standard Hamilton-Jacobi equation, you can reproduce the canonical quantization scheme.

The scheme was implemented for the case of a codimension one membrane (domain wall), since in this case the membrane dust is described using a single scalar function $\varphi$, for which the membranes are level surfaces $\varphi=\const$. This scalar function corresponds to the action variable in the Hamilton-Jacobi equation.

This scheme of quantization of the membrane wall naturally gives the nonlinear Klein-Fock-Gordon equation
$$
\left(|\psi|^{-4}+\Box\right)\psi=0.
$$

This approach to membrane quantization differs from the standard approach adopted in string theory. Therefore, quantization of the same systems can provide different results. In these cases, the discrepancy with the generally accepted results is not a disadvantage, since we consider a different physical model.

\section{Classical domain wall}

\subsection{Single domain wall action}

The action for a single domain wall is a standard Nambu-Goto type action for a relativistic membrane.
It is a measure of the world surface defined by an induced metric $h_{\alpha\beta}$.
Fields $X^M(\xi^\alpha)$ are space-time coordinates $X^M$ defined as functions of coordinates on the world surface $\xi^\alpha$.
\be\label{S0}
  S_0[X^M(\xi^\alpha)]=-T\int \sqrt{-\det h_{\alpha\beta}}\,d^{D-1}\xi,
\ee
$$
  M=0,\dots D-1,\quad\alpha,\beta=1,\dots,D-1,
$$
$$
  h_{\alpha\beta}=g_{MN}\frac{\p X^M}{\p\xi^\alpha}\frac{\p X^N}{\p\xi^\beta},
$$
where $g_{MN}(X)$ is the space-time metric with signature $(-,+,+,\dots,+)$.

\subsection{Domain wall dust action 1}

Let there be a family of non-interacting domain walls numbered by the continuous parameter $\phi$, then the corresponding action differs from \eqref{S0} by integrating over the parameter $\phi$
\be\label{S1}
  S_1[X^M(\xi^\alpha,\phi)]=-\int \sqrt{-\det h_{\alpha\beta}}\,d^{D-1}\xi\,d\phi,
\ee
$$
  M=0,\dots D-1,\quad\alpha,\beta=1,\dots,D-1
$$
$$
  h_{\alpha\beta}=g_{MN}\frac{\p X^M}{\p\xi^\alpha}\frac{\p X^N}{\p\xi^\beta},
$$
where $X^M$ are the Euler space-time coordinates.

Let the world surfaces of the domain walls corresponding to different values $\phi$ not intersect, and the Jacobian $\frac{DX}{D(\xi,\phi)}\not=0$, then
$\xi^\alpha$ and $\phi$ can also be considered as space-time coordinates, which are naturally 
called Lagrangian coordinates.

The functions $X^M(\xi,\phi)$ (dynamic fields) represent the Euler coordinates as functions of the Lagrangian coordinates.
$(\xi,\phi)\to X$.

\subsection{Domain wall dust action 2 (to build the Hamilton-Jakobi type equation)}

$X^M$ are the Euler space-time coordinates.

$\xi^\alpha$ and $\phi$ are the Lagrangian space-time coordinates.

Previously, \eqref{S1} we considered fields to be the Euler coordinates as functions of the Lagrangian coordinates, now we will take as fields inverse functions \cite{mgi-dan}, \cite{mgi-GnC}.

$\varphi(X^M)$ and $\xi^\alpha(X^M)$ are new dynamic fields. $X\to(\xi,\varphi)$.
(Here, $\phi$ is the independent variable, meanwhile $\varphi$ is the same coordinate, represented as function of $X$.)
\bea
  S_1[X^M(\xi^\alpha,\phi)]&=&-\int \sqrt{-\det h_{\alpha\beta}}\,d^{D-1}\xi\,d\phi=\nonumber\\
  &=&-\int \sqrt{g^{MN}\frac{\p\varphi}{\p X^M}\frac{\p\varphi}{\p X^N}}\,\sqrt{-g}\,d^DX=S_2[\varphi(X^M)],
  \label{S2}
\eea
$$
  g=\det g_{MN}.
$$

\textbf{The new action does not depend on the fields $\xi^\alpha(X)$!}

The domain walls are defined as level surfaces of the field $\varphi(X)$
$$
  \varphi(X)=\const.
$$

\subsection{Field equation and energy-momentum tensor}

The equation of field $\varphi$ is equivalent to the standard membrane equation of motion at all world surfaces
$\varphi=\const$
\be\label{fe-varphi}
  \frac1{\sqrt{-g}}\frac{\delta S_2}{\delta\varphi}
  =\underbrace{\frac{1}{\sqrt{-g}}\frac{\p~}{\p X^M}\sqrt{-g}}_{\Div}\,
  \underbrace{\frac{g^{MN}\frac{\p\varphi}{\p X^N}}{\sqrt{g^{KL}\frac{\p\varphi}{\p X^K}\frac{\p\varphi}{\p X^L}}}}_{\frac{\grad\varphi}{\|\grad\varphi\|}}=0.
\ee

The equivalence of the actions \eqref{S1} and \eqref{S2} (provided that the mapping $X\to (\xi,\varphi)$ is non-degenerate) is obvious from the form of the energy-momentum tensor (the field equations can be derived from the continuity equations $\nabla_MT^{MN}=0$).

The energy-momentum tensor has the form of a scalar multiplied by the orthogonal projector $P_{MN}$ on the surface
 $\varphi=\const$
\be\label{T_MN}
  T^{MN}=\frac{2}{\sqrt{-g}}\frac{\delta S_2}{\delta g_{MN}}=-\sqrt{g^{KL}\frac{\p\varphi}{\p X^K}\frac{\p\varphi}{\p X^L}}\,P^{MN},
\ee
$$
  P_{MN}=g_{MN}-\frac{\frac{\p\varphi}{\p X^M}\frac{\p\varphi}{\p X^N}}{g^{KL}\frac{\p\varphi}{\p X^K}\frac{\p\varphi}{\p X^L}},
$$
$$
  P^M_N P^N_K=P^M_K,\qquad P_{MN}=P_{NM},\qquad P^M_M=D-1.
$$

\subsection{Perturbation and causality}

It is easy to consider linear perturbations for a trivial solution of the form $\varphi_0=c z$ ($z=x^{D-1}$, $c=\const$) in Minkowski space.
However, as will be shown below, the general solution in arbitrary space-time is locally reduced to this case.

In a small area of space-time, we can choose coordinates in which the metric tensor has the form of a Minkowski metric, with the first derivatives of the metric (and the connectivity coefficients) equal to zero.

After that, using transformations from the Lorentz group, we can ensure that the $\varphi$ field is locally equal in the linear order (up to the $c=\const$ multiplier) to one of the spatial coordinates, which we denote $z$,
$$
  \varphi_0=cz+o(X).
$$
In the quadratic order in the coordinates
\be\label{varphi0}
  \varphi_0=cz+\frac{c}{2}(Kz^2+2K_mX^mz+K_{mn}X^mX^n)+o(X^2),\quad m,n=0,1\dots,D-2,~~ z=X^{D-1},
\ee
Here, $K,K_m,K_{mn}=\const$, $\p_M=\frac{\p~}{\p X^M}$.
$$
  \p_M\varphi_0=\left(\begin{array}{c}
  \p_m\varphi_0\\
  \p_z\varphi_0
  \end{array}\right)
  =c\left(\begin{array}{c}
  K_mz+K_{mn}X^n\\
  1+Kz+K_mX^m
  \end{array}\right)+o(X),
$$
$$
  (\p_M\varphi_0)(\p^M\varphi_0)=c^2(1+2Kz+2K_mX^m)+o(X),
$$
$$
  \frac{\p_M\varphi_0}{\sqrt{(\p_N\varphi_0)(\p^N\varphi_0)}}
  =\left(\begin{array}{c}
    K_mz+K_{mn}X^n\\
    1+Kz+K_mX^m
    \end{array}\right)(1-Kz-K_mX^m)+o(X)
  =\left(\begin{array}{c}
      K_mz+K_{mn}X^n\\
      1
   \end{array}\right)+o(X).
$$
Substituting the resulting expression into the field equation, we find a condition for the expansion coefficients (indeces are raised and lowered using the Minkowski metric)
\be\label{Kmm}
  \left.\p_M\frac{\p^M\varphi_0}{\sqrt{(\p_N\varphi_0)(\p^N\varphi_0)}}\right|_{X=0}=K^m_m=0.
\ee

We will consider the $\varphi_0$ field as an unperturbed solution.
  
Now let us add a small perturbation to the field
$$
  \varphi(X)=\varphi_0(X)+\varepsilon cf(X),\qquad \varepsilon=\const\ll1.
$$

$$
  \p_M\varphi=c\left(\begin{array}{c}
  K_mz+K_{mn}X^n+\varepsilon\p_mf\\
  1+Kz+K_mX^m+\varepsilon\p_zf
  \end{array}\right)+o(X)+o(\varepsilon),
$$
$$
  (\p_M\varphi)(\p^M\varphi)=c^2(1+2Kz+2K_mX^m+2\varepsilon\p_zf)+o(X)+o(\varepsilon),
$$
\bean
  \frac{\p_M\varphi}{\sqrt{(\p_N\varphi)(\p^N\varphi)}}
  &=&\left(\begin{array}{c}
    K_mz+K_{mn}X^n+\varepsilon\p_mf\\
    1+Kz+K_mX^m+\varepsilon\p_zf
    \end{array}\right)(1-Kz-K_mX^m-\varepsilon\p_zf)+o(X)+o(\varepsilon)=\\
  &=&\left(\begin{array}{c}
      K_mz+K_{mn}X^n+\varepsilon\p_mf\\
      1
   \end{array}\right)+o(X)+o(\varepsilon).
\eean
\be\label{Kmm2}
  \left.\p_M\frac{\p^M\varphi}{\sqrt{(\p_N\varphi)(\p^N\varphi_0)}}\right|_{X=0}
  =\underbrace{K^m_m}_0+\varepsilon\p_m\p^mf+o(\varepsilon)=0.
\ee
Thus for a linear perturbation, we have a wave equation on the world surface of the membrane wall
$$
  \p_m\p^mf=0,\qquad m=0,1,\dots,D-2.
$$
The perturbation is transferred along the membrane wall at a unit speed (i.e., the speed of light), which means that the principle of causality is valid.

\subsection{The Hamilton-Jacobi type equation}

The field equation for domain wall dust is somewhat similar to the Hamilton-Jacobi equation
$$
  \Div \frac{\grad\varphi}{\|\grad\varphi\|}=0.
$$
It describes a set of non-overlapping non-interacting domain walls with different initial conditions.

\textbf{However, it is a second-order equation, whereas the Hamilton-Jacobi equation is always first-order.}

Let us introduce an additional field
\be\label{HJ}
  \rho=\frac{1}{\|\grad\varphi\|}.
\ee
The equation linking $\rho$ and $\phi$ is a first-order differential equation on $\phi$, and it can be considered as a Hamilton-Jacobi type equation (the Hamilton-Jacobi equation depending on the functional parameter $\rho$).

The original equation of the field \eqref{fe-varphi}, rewritten through the field $\rho$, takes the form of a continuity equation
\be\label{div-rho}
  \Div(\rho\,\grad\varphi)=0.
\ee

The equations \eqref{HJ}, \eqref{div-rho} are nontrivial.

First, the continuity equation \eqref{div-rho} \textit{looks} tachyonic ($\grad\varphi$ is space-like), but the perturbations are not tachyonic, this is evident from the absence of tachyons for the domain wall 
(perturbations for an elastic medium similarly described via the Lagrangian coordinate defined as functions of the Euler coordinates are considered in \cite{perturb}).
$$
  \Div(\rho\,\grad\varphi)=\underbrace{\frac{1}{\sqrt{g}}\frac{\p~}{\p X^M}\sqrt{g}}_{\Div}
  \,\rho\, \underbrace{g^{MN}\frac{\p\varphi}{\p X^N}}_{\grad\varphi}=0,
$$

Second, the Hamilton-Jacobi type equation \eqref{HJ} depends on the density $\rho$
$$
  \|\grad\varphi\|^2=g^{KL}\frac{\p\varphi}{\p X^K}\frac{\p\varphi}{\p X^L}=\frac{1}{\rho^2}.
$$

\subsection{Domain wall dust action 3 (for quantization)}

It is easy to find an action for $\varphi$ and $\rho$ as independent fields that would reproduce the equations \eqref {HJ}, \eqref{div-rho} as the Euler-Lagrange equations.
\be\label{S3}
  S_2[\rho(X),\varphi(X)]=-\frac12\int\left(\rho\,\|\grad\varphi\|^2+\frac1\rho\right)\,\sqrt{-g}\,d^DX.
\ee

The Euler-Lagrange equations reproduce \eqref{HJ}, \eqref{div-rho}.
$$
  \frac{1}{\sqrt{-g}}\frac{\delta S_2}{\delta\rho}=\frac12\left(\frac{1}{\rho^2}-\|\grad\varphi\|^2\right)=0,
$$
$$
  \frac{1}{\sqrt{-g}}\frac{\delta S_2}{\delta\varphi}=\frac12\Div(\rho\,\grad\varphi)=0,
$$
$$
  T_{MN}=\rho\frac{\p\varphi}{\p X^M}\frac{\p\varphi}{\p X^N}
  -\frac12\,g_{MN}\,\rho\,\left(\|\grad\varphi\|^2+\frac{1}{\rho^2}\right),
$$

On solutions of the field equations (if we impose the constraint \eqref{HJ} between the $\varphi$ and $\rho$ fields), the energy-momentum tensor coincides with the previously obtained \eqref{T_MN}
$$
  \left.T_{MN}\right|_{\frac{\delta S}{\delta\rho}=0}=-\|\grad\varphi\| P_{MN},
\qquad
  P_{MN}=g_{MN}-\frac{\frac{\p\varphi}{\p X^M}\frac{\p\varphi}{\p X^N}}{g^{KL}\frac{\p\varphi}{\p X^K}\frac{\p\varphi}{\p X^L}}.
$$

\section{Quantization}

\subsection{Preparing for quantization}

The Hamilton-Jacobi type equation
$$
  \frac{1}{\sqrt{-g}}\frac{\delta S_2}{\delta\rho}=\frac12\left(\frac{1}{\rho^2}-\|\grad\varphi\|^2\right)=0
$$
allows one to find the ``extended Hamiltonian'' (see, for example, \cite{exHam}) by substitution
$\frac{\p\varphi}{\p X^M}\to P_M$.
$$
  H(X^M,P_M)=\frac12\left(\frac{1}{\rho^2(X)}-P_MP^M\right).
$$
The extended Hamiltonian includes a time component of the relativistic impulse.
On the energy surface it vanishes.

The extended Hamiltonian depends on $\rho(X)$.

\subsection{Canonical quantization}
Let us replace the momenta in the extended Hamiltonian with the corresponding operators, 
and express the density $\rho$ in terms of $\psi$
$$
  P_M\to \hat P_M=-\I\frac{\p~}{\p X^M},\qquad \rho(X)=|\psi(X)|^2.
$$
$$
  H(X^M,P_M)=\frac12\left(\frac{1}{\rho^2(X)}-P_MP^M\right)\to
  \hat H[\psi]=\frac12\left(\frac{1}{|\psi(X)|^4}-\hat P_M\hat P^M\right)
$$

The corresponding equation is the nonlinear equation of Klein-Fock-Gordon
$$
  \hat H[\psi]\psi=0
$$
$$
  \frac12\left(\frac{1}{|\psi(X)|^4}+\Box\right)\psi(X)=0
$$
$$
  \Box=\frac{1}{\sqrt{-g}}\frac{\p~}{\p X^M}\sqrt{-g}\,g^{MN}\frac{\p~}{\p X^N}.
$$

\subsection{Quantum action}
We can reproduce the nonlinear Klein-Fock-Gordon equation using the following action functional ($\psi$ and $\I\psi^*$ are the canonical variables)
$$
  \frac12\left(\frac{1}{|\psi(X)|^4}+\Box\right)\psi(X)=0
$$
$$
  S_q[\psi(X),\mathrm{i}\psi^*(X)]=-\frac12\int \left(\frac{1}{\psi^*\psi}
  +g^{MN}\frac{\p\psi^*}{\p X^M}\frac{\p\psi}{\p X^N}\right)\,\sqrt{-g}\,d^DX
$$
$$
  \frac{1}{\sqrt{-g}}\frac{\delta S_q}{\delta\psi^*}=\frac12\left(\frac{1}{|\psi(X)|^4}+\Box\right)\psi(X)=0,
$$
$$
  \frac{1}{\sqrt{-g}}\frac{\delta S_q}{\delta\psi}=\frac12\left(\frac{1}{|\psi(X)|^4}+\Box\right)\psi^*(X)=0,
$$

\section{Quantum corrections}

To compare the resulting field equations with the classical case, we rewrite the action using the real fields $\rho$ and $\varphi$
$$
  \psi(X)=\sqrt{\rho(X)}\,\E^{\I\varphi(X)}.
$$
$$
  S_q[\psi(X),\mathrm{i}\psi^*(X)]=-\frac12\int \left(\frac{1}{\psi^*\psi}
  +g^{MN}\frac{\p\psi^*}{\p X^M}\frac{\p\psi}{\p X^N}\right)\,\sqrt{-g}\,d^DX=
$$
$$
  =S_q[\rho(X),\varphi(X)]=-\frac12\int\Big(\frac{1}{\rho}+\rho\,\|\grad\varphi\|^2
  +\underbrace{\|\grad\sqrt{\rho}\|^2}_\text{q.corrction}
  \Big)\,\sqrt{-g}\,d^DX.
$$
By the variational derivatives we find
$$
  \frac{1}{\sqrt{-g}}\frac{\delta S_q}{\delta\rho}=\frac12\Big(\frac{1}{\rho^2}-\|\grad\varphi\|^2
  +\underbrace{\frac{\Box\sqrt{\rho}}{\sqrt{\rho}}}_\text{q.corrction}\Big)=0
$$
$$
  \frac{1}{\sqrt{-g}}\frac{\delta S_q}{\delta\varphi}=\frac12\Div(\rho\,\grad\varphi)=0.
$$
\bean
  T_{MN}&=&\rho\frac{\p\varphi}{\p X^M}\frac{\p\varphi}{\p X^N}
  -\frac12\,g_{MN}\,\rho\,\left(\|\grad\varphi\|^2+\frac{1}{\rho^2}\right)+\\
  &&+\underbrace{\frac{\p\sqrt{\rho}}{\p X^M}\frac{\p\sqrt{\rho}}{\p X^N}  
  -\frac12\,g_{MN}\|\grad\sqrt{\rho}\|^2}_\text{q.correction}
\eean

Under the field equation $\frac{\delta S}{\delta\rho}=0$ energy-momentum tensor has the form
\bean
  \left.T_{MN}\right|_{\frac{\delta S}{\delta\rho}=0}&=&-\|\grad\varphi\|\,P_{MN}
  +\frac12\,g_{MN}\,\sqrt{\rho}\,\Box\sqrt{\rho}+\\
  &&+\frac{\p\sqrt{\rho}}{\p X^M}\frac{\p\sqrt{\rho}}{\p X^N}  
  -\frac12\,g_{MN}\|\grad\sqrt{\rho}\|^2=\\
  &=&-\|\grad\varphi\|\,P_{MN}+\\
  &&+\underbrace{\frac{\p\sqrt{\rho}}{\p X^M}\frac{\p\sqrt{\rho}}{\p X^N} 
  +\frac12\,g_{MN}\,\rho\,\Div\frac{\grad\sqrt{\rho}}{\sqrt{\rho}}}_\text{q.corrction}.
\eean

\subsection{Perturbation and causality}

Let us consider the unperturbed solution of the field equation based on the unperturbed solution \eqref{varphi0} in the classical case.
\be
  \varphi_0=cz+\frac{c}{2}(Kz^2+2K_mX^mz+K_{mn}X^mX^n)+o(X^2),~m,n=0,1\dots,D-2,~~ z=X^{D-1},~K^m_m=0,
\ee
Here, $K,K_m,K_{mn}=\const$.
\be
  \sqrt{\rho_0}=\frac{1}{\sqrt{c}}\left(1-\frac{Kz}{2}-\frac{K_mX^m}{2}\right)+o(X^2).
\ee
Let us look for a solution in the form of
$$
  \varphi(X)=\varphi_0(X)+\varepsilon c f(X)+o(X^2)+o(\varepsilon),\quad \varepsilon=\const\ll 1.
$$
$$
  \sqrt{\rho(X)}=\sqrt{\rho_0(X)}-\varepsilon\frac{g(X)}{2\sqrt{c}}+o(X^2)+o(\varepsilon).
$$
\bean
  \rho\p_M\varphi&=&\left(\begin{array}{c}
  K_mz+K_{mn}X^n+\varepsilon\p_mf\\
  1+Kz+K_mX^m+\varepsilon\p_zf
  \end{array}\right)(1-Kz-K_mX^m-\varepsilon g)+o(X)+o(\varepsilon)=\\
  &=&\left(\begin{array}{c}
    K_mz+K_{mn}X^n+\varepsilon\p_mf\\
    1+\varepsilon\p_zf-\varepsilon g
    \end{array}\right)+o(X)+o(\varepsilon).
\eean
$$
  \Div(\rho\,\grad\varphi)|_{X=0}=\varepsilon\left(\p_m\p^m f+\p_z^2 f-\p_z g\right)+o(\varepsilon)=0.
$$
Let us denote $g=\p_z f+h$, then we obtain the first equation for the perturbation
\be\label{f-q}
  \p_m\p^m f-\p_z h=0.
\ee
$$
  \rho^{-2}=c^2(1+2Kz+2K_mX^m+2\varepsilon(\p_z f+h))+o(X)+o(\varepsilon).
$$
$$
  \|\grad\varphi\|^2=c^2(1+2Kz+2K_mX^m+2\varepsilon\p_z f)+o(X)+o(\varepsilon).
$$
\bean
  \left.\frac{\Box\sqrt{\rho}}{\sqrt{\rho}}\right|_{X=0}
  &=&\left(1+\frac{Kz}{2}+\frac{K_mX^m}{2}+\frac{\varepsilon}{2}(\p_z f+h)\right)
  \Box\left(-\frac{\varepsilon}{2}(\p_z f+h)\right)+o(\varepsilon)=\\
  &=&-\frac{\varepsilon}{2}\Box(\p_zf+h)+o(\varepsilon).
\eean

In the classical limit $\rho^{-2}-\|\grad\varphi\|^2=0$, and we obtain $h=0$.
The equation \eqref{f-q} gives, as previously, $\p_m\p^m f=0$.

In the quantum case
\be\label{h-q}
  \frac{1}{\rho^2}-\|\grad\varphi\|^2+\frac{\Box\sqrt{\rho}}{\sqrt{\rho}}
  =\frac{\varepsilon}{2}\left[4c^2 h-\Box(\p_z f+h)\right]+o(\varepsilon)=0
\ee

Let 
\bea
  f&=&a\,\E^{\I P_M X^M},\\
  h&=&\I b\,\E^{\I P_M X^M},
\eea
The system \eqref{f-q}, \eqref{h-q} takes the form
\bea
  &&-P_mP^m a+P_z b=0,\\
  &&4c^2b+P_MP^M(P_za+b)=0.
\eea
$$
  -\omega^2+P_\mu^2+P_z^2=P_MP^M=-2c^2\left(1\pm\sqrt{1+\frac{P_z^2}{c^2}}\right)
$$
Hereinafter, $\mu=1,\dots,D-2$.

We obtained the dispersion relations
$$
  \omega=\sqrt{P_\mu^2+P_z^2+2c^2\left(1\pm\sqrt{1+\frac{P_z^2}{c^2}}\right)}
$$

The components of the group velocity are
$$
  v_\mu=\frac{\p\omega}{\p P_\mu}=\frac{P_\mu}{\omega},\qquad
  v_z=\frac{P_z}{\omega}\left(1\pm\frac{1}{\sqrt{1+\frac{P_z^2}{c^2}}}\right).
$$
It is easy to check that when selecting the upper sign, the group speed is strictly less than 1 (the speed of light),
and when selecting the lower sign, the group speed is less than 1 for non-zero values of $P_z$ and it turns to 1 for $P_z=0$.

Such perturbations, as in the classical case, do not violate the causality principle.

If one sets $P_z=0$ or, equivalently, $\p_zf=\p_zh=0$, then the equations \eqref{f-q}, \eqref{h-q} give
\bea
   &&\p_m\p^m f=0,\\
   &&(4c^2-\p_m\p^m)h=0.
\eea
In this case, the perturbations of the $f$ and $h$ fields propagate along the surface of the domain wall, and the $f$ field still has massless excitations, and the $h$ field has excitations with a mass depending on the density of the domain walls $m_h=2c>0$.

Stability of perturbations is an interesting problem.
We can expect that in case of increasing perturbations the system will switch to the classic mode, and the interaction between the domain walls will be turned off.
Thus the instability of small perturbations, if any, should not deprive small perturbations of physical meaning.

In any case, here we study causality, and the problem of stability in this context is insignificant.



\section{Conclusion}

The resulting equation has the form of a nonlinear Klein-Fock-Gordon equation.

Why is it nonlinear?
The most natural options are the following:
\begin{itemize}
\item Fundamental nonlinearity,
\item Mean field approximation,
\begin{itemize}
\item Mean field of membranes (domain walls),
\item or mean field of tachions ($\grad\varphi$ initially was space-like),
\item Is domain wall a sort of tachyonic condansate?
\end{itemize}
\end{itemize}

The result may be interesting in membrane quantization in superstring and supergravity theories \cite{obzor} and in condensed matter physics \cite{perturb}.

It is interesting to consider nonlinear versions of quantum field theory with renormalizations based on the representation of coordinates and momenta as a set of discrete variables (digits in the positional number system).
Renormalizations based on the binary number system were introduced in \cite{mgi-binary}.

\end{document}